\documentclass[12pt]{article}
 \newcommand{\be}{\begin{equation}}
 \newcommand{\ee}{\end{equation}}
 \newcommand{\ba}{\begin{eqnarray}}
 \newcommand{\ea}{\end{eqnarray}}
 \def\Journal#1#2#3#4{{#1} {#2}, #3 (#4)}

 \def\ASP{Astropart.\ Physics}
 \def\CQG{Class.\ Quantum Grav.}
 \def\GRG{Gen.\ Rel. Grav.}

 \def\NPB{Nucl.\ Phys.\ B}
 
 \def\PLB{Phys.\ Lett.\  B}
 \def\PRL{Phys.\ Rev.\ Lett.}
 \def\PRD{Phys.\ Rev.\ D}
 \def\PRP{Phys.\ Rep.}
 
 \def\laq{\,\raise 0.4ex\hbox{$<$}\kern -0.8em\lower 0.62ex\hbox{$\sim$}\,}
 \def\gaq{\,\raise 0.4ex\hbox{$>$}\kern -0.7em\lower 0.62ex\hbox{$\sim$}\,}

\begin{document}

\title{FRW Cosmological Solutions in M-theory}

\author{Marco Cavagli\`a~$^{(A,B,C)}$,  
\underline{Paulo Vargas Moniz}~$^{(B,D)}$\\
\small $^{(A)}$ Center for Theoretical Physics, Massachusetts Institute for
Technology\\ 
\small 77 Massachusetts Avenue, Cambridge MA 02139, USA\\
\small $^{(B)}$ Departamento de F{\'\i}sica,
Universidade da Beira Interior,\\
\small Rua Marqu{\^e}s d'{\'A}vila e Bolama,
6200 Covilh{\~a}, Portugal\\
\small $^{(C)}$ INFN, Sede di Presidenza, Roma, Italia\\
\small $^{(D)}$ CENTRA, IST, Av. Rovisco Pais, 1049 
Lisboa Codex, Portugal}
\date{(MIT-CTP-3050, gr-qc/0011098. November 27, 2000)}
\maketitle
\abstract{
We present the canonical and quantum cosmological investigation of a
four-di\-men\-sio\-nal, spatially flat, Friedmann-Robertson-Walker (FRW) model
that is derived from the bosonic Neveu-Schwarz/Neveu-Schwarz sector of the
low-energy M-theory effective action. We discuss in detail the phase space of
the classical theory. We find the quantum solutions of the model and obtain the
positive norm Hilbert space of states. Finally, the correspondence between wave
functions and classical solutions is outlined.

}
\section{Introduction}
The search for a theory of quantum gravity constitutes one of the foremost
challenges in theoretical high energy physics. The need for quantum gravity
finds its roots within Einstein general relativity. Powerful general theorems
imply that our universe must have started from an initially singular state with
infinite curvature. In such circunstances, where the laws of classical physics
break down, it is unclear how any boundary conditions necessary for a
description of a dynamical system could have been imposed at the initial
singularity. Quantum corrections could then induce a modification of classical
general relativity and strongly influence the evolution of the very early
universe.

In the last two decades superstring theory\cite{strings} has emerged as a
successful candidate for the theory of quantum gravity. In cosmology, most of
the modifications to general relativity induced by superstring theory are
originated by the inclusion of the dilaton, axion and various moduli fields,
together with higher curvature terms that are present in the low-energy
effective actions. Each of these novel ingredients leads to new cosmological
solutions. A remarkable example is given by the so-called pre-big bang
scenario\cite{prebb} that follows from the low-energy effective string action.
Different branches of the solution are related by time reflection and internal
transformations -- $O(d,d)$ and, in particular, scale factor duality --  that
descend from the $T$-duality property of the full superstring theory. According
to the pre-big bang scenario the universe evolves from a weak-coupled string
vacuum state to a radiation-dominated and a subsequent matter-dominated FRW
geometry going through a region of strong coupling and large curvature.
Although the pre-big bang model has not yet proven able to solve all its
pitfalls, such as the existence of a singular boundary that separates the pre-
and post-big bang branches\cite{exit}, it provides a good starting point to
investigate high-energy cosmology.

Recently, it has been argued that the five consistent, anomaly free,
perturbative formulations of ten-dimensional superstring theories are connected
by a web of duality transformations and constitute special points of a large,
multi-dimensional moduli space of a fundamental (non-perturbative) theory,
called M-theory. Quite interestingly, another point of the moduli space of
M-theory corresponds to eleven-dimensional supergravity, which is the
low-energy limit of M-theory. Assuming that M-theory is the ultimate theory of
quantum gravity, it is natural to explore its cosmological implications.
Although our understanding of M-theory is still incomplete, there are hopes
that some of the obstacles of dilaton driven inflation in string theory could
be overcome within the new theory. The underlying idea is to investigate the
dynamics at the extreme weak- and strong-coupling regimes of superstring theory
from a M-theory perspective, where the existence of eleven dimensions seems
mandatory. 

Several approaches to M-theory cosmology have been explored in the
literature\cite{LOW,RS,DH}-\cite{BCLN}. In the framework of the Ho{\v
r}ava-Witten model\cite{HW}, M-theory and cosmology have been combined in the
works of Lukas, Ovrut and Waldram\cite{LOW}. A somewhat related line of
research is the brane world by Randall and Sundrum\cite{RS}, where our
four-dimensional universe emerges as the world volume of a three-brane in a
higher-dimensional spacetime. From a different point of view, Damour and
Henneaux have investigated chaotic models\cite{DH} and Lu, Maharana, Mukherji
and Pope have discussed classical and quantum M-theory models with homogeneous
graviton, dilaton and antisymmetric tensor field strengths\cite{LMMP}.
Different classes of cosmological solutions that reduce to solutions of string
dilaton gravity have been discussed\cite{KKO}. In particular, a global analysis
of four-dimensional cosmologies derived from M-theory and type $IIA$
superstring theory has been presented by Billyard, Coley, Lidsey and Nilsson
(BCLN)\cite{BCLN}. Using the theory of dynamical systems to determine the
qualitative behaviour of the solutions, the authors find that fields associated
with the Neveu/Schwarz-Neveu/Schwarz (NSNS) and Ramond-Ramond (RR) sectors play
a rather crucial role in determining the dynamical behaviour of the solutions.
Quite interestingly, for spatially flat FRW models the boundary of the
classical physical phase space is a set of invariant submanifolds, where either
the axion field is trivial or the RR four-form field strength is dynamically
unimportant. This interplay leads to important consequences, as the orbits in
the phase space are dominated by the dynamics associated with one, or the
other, or both invariant submanifolds in sequence, shadowing trajectories in
the invariant submanifold\cite{BCLN}. 

In this talk we discuss the main scenario introduced by BCLN from a canonical
perspective. This approach allows us to analyse in depth the physical
properties of the classical solutions and to obtain a consistent quantum
description of the model. We consider the bosonic sector of eleven-dimensional
supergravity which consists of a graviton and an antisymmetric three-form
potential. The theory is compactified to four dimensions by assuming a geometry
of the form $M^4 \times T^6 \times S^1$, where $T^6$ is a six-dimensional torus
and $M^4$ corresponds to a spatially flat FRW spacetime. The effective theory
in four dimensions bears a dilaton $\phi$, a modulus field $\beta$ identifying
the internal space, a pseudo-scalar axion field $\sigma$ and a potential term
induced by the RR four-form field.  A brief derivation of the previous steps is
presented in Section 2. In Section 3 we analyse the NSNS model, where the
four-form field is negligible and the axion field dominates. In particular, in
Subsection 3.1 we discuss the parameter space of the classical theory and in
Subsection 3.2 we find the Hilbert space of states of the quantum theory. This
programme is performed using a set of canonical variables, the so-called
``hybrid" variables, that diagonalise the Hamiltonian. More details will
appear in a forthcoming report\cite{MC-PM}. Finally, our conclusions are
presented in Section 4. 
\section{M-theory cosmology\label{cosmology}}
In this section we derive\cite{BCLN} the four-dimensional minisuperspace
effective action that will be used to discuss the dynamics of M-theory
cosmology. 

The bosonic sector of eleven-dimensional supergravity action $S^{(11)}$ is
\ba
&&S^{(11)}=\int d^{11}X \sqrt{-g^{(11)}}\left[R^{(11)}(g^{(11)}_{ab})
-{1\over 48}F_{a_1\dots a_4}F^{a_1\dots a_4}\right.\nonumber\\
&&\qquad\qquad\left. -{1\over 12^4\sqrt{-g^{(11)}}}\epsilon^{a_1 \dots a_3 
b_1 \dots b_4 c_1 \dots c_4}
A_{a_1 \dots a_3}F_{b_1 \dots b_4}F_{c_1 \dots c_4} \right]\,,
\label{b1}\ea
where $a_i,b_i,c_i=0\dots 10$, $F_{a_1 \dots a_4}=4\partial_{[a_1}A_{a_2 \dots
a_4]}$ is the four-form field strength of the antisymmetric three-form
potential $A_{a_1 \dots a_3}$, and $g^{(11)}$ denotes the determinant of the
eleven-dimensional metric $g^{(11)}_{ab}$. Equation (\ref{b1}) describes the
low-energy limit of M-theory.

The four-dimensional effective action is derived from Eq.\ (\ref{b1}) by a
sequence of a Kaluza-Klein compactification on a circle $S^1$ with radius
$R_{S^1}=e^{\Phi_{10}/3}$, a conformal transformation of the ensuing
ten-dimensional metric with conformal factor $R_{S^1}^{-1}$, and a further
compactification on an isotropic six-torus with radius $R_{T^6}=e^{\beta}$. 

We are interested in homogeneous and isotropic four-dimensional
cosmologies. The ansatz for the four-dimensional section of the metric in the
string frame is
\be
ds^2_{(4)}\equiv
g_{\mu\nu}dx^\mu dx^\nu=-N^2(t)dt^2+e^{2\alpha(t)}d\Omega_{3k}\,,\qquad
N(t)>0
\label{c1}
\ee
where $d\Omega_{3k}$ is a maximally symmetric three-dimensional metric with
unit volume and curvature $k=0,\pm 1$, respectively. Using Eq.\ (\ref{c1})  and
requiring that the modulus field $\beta$, the dilaton $\Phi_4$, and the axion
$\sigma$ depend only on $t$, the four-dimensional effective action is
\be
S=\int
dt\left[{1\over\mu}\left(3\dot\alpha^2-\dot\phi^2
+6\dot\beta^2+{\dot\sigma^2\over 2~}e^{2(3\alpha+\phi)}\right)+
\mu\left(6k e^{-2(\alpha+\phi)}-{Q^2\over
2~}e^{3\alpha-\phi-6\beta}\right)\right],
\label{c2}
\ee
where we have defined the ``shifted dilaton'' field
\be
\phi=\Phi_4-3\alpha\,,
\label{c3}
\ee
and the Lagrange multiplier
\be
\mu(t)=Ne^{\phi}>0\,.\label{mu}
\label{c4}
\ee
The dynamics of the action (\ref{c2}) has been discussed qualitatively in Ref.\
\cite{BCLN}. Here we discuss in detail the model with $Q=0$ and $k=0$. This
case turns out to be completely integrable and describes spatially flat NSNS
low-energy M-theory cosmology with negligible RR fields. The general solution
for this model (including spatially curved models which are not discussed here)
was first discussed by Copeland, Lahiri and Wands\cite{CLW}. Let us note that
the model with constant $\sigma$ and $k=0$ is also completely integrable and
has been discussed quantitatively in Ref.\ \cite{MC-PM}. The latter case
describes spatially flat low-energy M-theory cosmology with trivial axion and
nonzero RR four-form. 
\section{NSNS low-energy M-theory cosmology}
The action can be cast in the canonical form
\be
S_I=\int dt \left[\dot\alpha p_\alpha+\dot\phi p_\phi+\dot\beta
p_\beta+\dot\sigma p_\sigma-{\cal H}\right]\,,
\label{d1}
\ee
where the Hamiltonian is
\be
{\cal H}=\mu H\,,\qquad
H={1\over 24}\left(2p^2_\alpha-6p^2_\phi+p^2_\beta
+12p^2_\sigma e^{-2(3\alpha+\phi)}\right)\,.\label{HII}
\label{d2}
\ee
The non-dynamical variable $\mu$ enforces the Hamiltonian constraint 
\be
0=24H=2p_\alpha^2-6p^2_\phi+p^2_\beta+12p^2_\sigma e^{-2(3\alpha+\phi)}\,.
\label{constraintII}
\ee
The canonical equations of motion are
\be
\begin{array}{llll}
\displaystyle
\dot\alpha={p_\alpha\over 6}\,,\quad
&\displaystyle
\dot\phi=-{p_\phi\over 2}\,,\quad
&\displaystyle
\dot\beta={p_\beta\over 12}\,,\quad
&\displaystyle
\dot\sigma=p_\sigma e^{-2(3\alpha+\phi)}\,,\\\\
\displaystyle
\dot p_\alpha=3p_\sigma^2e^{-2(3\alpha+\phi)}\,,\quad
&\displaystyle
\dot p_\phi=p_\sigma^2e^{-2(3\alpha+\phi)}\,,\quad
&\displaystyle
\dot p_\beta=0\,,\quad
&\displaystyle
\dot p_\sigma=0\,,
\end{array}\label{eqsII}
\ee
where the dots represent differentiation w.r.t.\ gauge parameter
\be
\tau(t)=\int_{t_0}^t \mu(t') dt'\,,\qquad t>t_0\,,
\label{d5}
\ee
and $t_0$ is an arbitrary constant. Note that since $\mu$ is positive defined
$\tau(t)$ is a monotonic increasing function.
\subsection{Classical solutions}
Assuming $p_\sigma\not=0$ the off-shell solution of the canonical equations
is
\be
\begin{array}{lll}
\alpha&=&\displaystyle\alpha_0+{1\over 2}\ln
\left[\cosh\left(\kappa(\tau-\tau_0)\right)\right]
-\xi(\tau-\tau_0)\,,\\\\
p_\alpha&=&\displaystyle 3\kappa\tanh\left[\kappa(\tau-\tau_0)\right]-6\xi\,,\\\\
\phi&=&\displaystyle\phi_0-{1\over 2}\ln
\left[\cosh\left(\kappa(\tau-\tau_0)\right)\right]+
3\xi(\tau-\tau_0)\,,\label{phi}\\\nonumber\\
p_\phi&=&\displaystyle\kappa\tanh\left[\kappa(\tau-\tau_0)\right]-6\xi\,,\\\\
\beta&=&\displaystyle\beta_0+{p_\beta\over 12}(\tau-\tau_0)\,,\\\\
p_\beta&=&{\rm constant}\,,\\\\
\sigma&=&\displaystyle\sigma_0+{\kappa\over
p_\sigma}\tanh\left[\kappa(\tau-\tau_0)\right]\\\\
p_\sigma&=&{\rm constant}\,,
\end{array}
\label{solII}
\ee
where $\alpha_0$, $\phi_0$, $\beta_0$, $\sigma_0$ and $\tau_0$ are constants of
integration, 
\be
\kappa^2-12\xi^2+{p_\beta^2\over 12}=2H\,,\qquad\kappa\not=0\,,
\label{kappaII}
\ee
and (we choose $\kappa>0$ for simplicity)
\be
3\alpha_0+\phi_0=\ln\left({|p_\sigma|\over\kappa}\right)\,.
\ee
A useful canonical chart is formed by the hybrid variables that
diagonalise the constraint (\ref{constraintII}). Although the hybrid variables
are not (all) gauge invariant they allow to fix a global gauge and quantize
exactly the system. The hybrid variables $(a,b,c,\sigma)$ are defined by the
canonical transformation
\be
\displaystyle
\begin{array}{lll}
\displaystyle
a=\phi+3\alpha\,,\quad
&\displaystyle
b=\sqrt{3}(\phi+\alpha)\,,\quad
&\displaystyle
c=2\sqrt{3}\beta\,,\\\\
\displaystyle
p_a={1\over 2}\left(p_\alpha-p_\phi\right)\,,\quad
&\displaystyle
p_b={1\over 2\sqrt{3}}\left(3p_\phi-p_\alpha\right)\,,\quad
&\displaystyle
p_c={1\over 2\sqrt{3}}p_\beta\,.
\end{array}\label{hybridII}
\ee
Note that $a$ coincides with the four-dimensional dilaton field $\Phi_4$. Using
the hybrid variables the constraint (\ref{constraintII}) reads (we have divided
by a factor $12$)
\be
p_a^2-p^2_b+p^2_c+p^2_\sigma e^{-2a}=0\,.
\label{constraint-hybII}
\ee

Let us discuss the behaviour of the classical solution (\ref{solII}). The
on-shell classical solution is determined by six physical parameters. Five of
them ($\alpha_0$, $\phi_0$, $\beta_0$, $\sigma_0$, and $\tau_0$) give initial
conditions for the canonical variables and will be set equal to zero.
Therefore, the qualitative behaviour of the model is determined by a
two-dimensional parameter space described by two coordinates, for instance
$\kappa$ and $\xi$. Using $\kappa$ and $\xi$ as free parameters, from Eq.\
(\ref{kappaII}) it follows that $p_\beta$ is (on-shell)
\be
p_\beta=\pm{2\sqrt{3}}\sqrt{12\xi^2-\kappa^2}\,.
\label{hypI}
\ee
The sign of $p_\beta$ determines the dynamical behaviour of the internal
six-torus space. From the solution of the equations of motion one obtains the
scale factor of the internal space 
\be
R_{T^6}=e^{p_\beta\tau/12}\,.
\ee
A successful physical model ultimately requires that the moduli fields are
stabilized and compactified at late times. Stabilization of the internal space
does not occur in the models under consideration, where only a fraction of all
the degrees of freedom present in Eq.\ (\ref{b1}) are considered, with
exception of the (fine-tuned) case $p_\beta=0$. (Hopefully, the inclusion of
more degrees of freedom will provide a mechanism for stabilization of
extra-dimensions at late times.) Compactification of the six-torus space is
achieved for $p_\beta>0$. Indeed, for negative values of $p_\beta$ the internal
space shrinks to zero for large values of the gauge time $\tau$ and
decompatifies for $\tau\to-\infty$ when the strong coupling region of the
theory is approached. Since the relation between the comoving ($N=1$) time $t$
and the gauge time is monotonic, the dynamics in $\tau$ traces the dynamics in
$t$, and the internal space shrinks to zero for large values of the (physical)
comoving time as well. The fine-tuned limiting value $p_\beta=0$ corresponds to
a constant (stable) internal space with unit radius. In the following we will
restrict attention to nonpositive values of $p_\beta$.

At fixed $\kappa$ we distinguish three different dynamical behaviours of the
four-dimensional external space according to the value of $\xi$:
\begin{itemize} 
\item[{\it i)}] $\xi\le-\kappa/2$ ($p_\beta<-2\kappa$). In this case the
external scale factor always expands while the internal scale factor shrinks from
infinity to zero. The Hubble parameter is always positive and vanishes
asymptotically at large times. In particular, for $\xi=-\kappa/2$ the external
space starts at $\tau=-\infty$ with a finite nonzero scale factor and vanishing
Hubble parameter. For $\xi<-\kappa/2$ the external space starts with a
vanishing scale factor and infinite Hubble parameter, which is always
decreasing. $\tau=-\infty$ is the strong coupling region where both the
coupling constants of the theory, $g=\exp(\phi)$ and $g_{10}=\exp(\Phi_{10})$,
become infinite. Conversely, $\tau=\infty$ is the weak region coupling where
$g$ and $g_{10}$ vanish. $g$ and $g_{10}$ are always decreasing.  
\item[{\it ii)}] $\xi\ge\kappa/2$ ($p_\beta<-2\kappa$). In this case both the
external scale factor and the internal scale factor always shrink. The Hubble
parameter is always negative and asymptotically vanishing at small times. In
particular, for $\xi=\kappa/2$ the external space ends at $\tau=\infty$ with a
finite nonzero scale factor and vanishing Hubble parameter. For $\xi>\kappa/2$
the external space ends with a vanishing scale factor and infinite Hubble
parameter. $g$ ($g_{10}$) increases (decreases) from zero (infinity) to
infinity (zero). 
\item[{\it iii)}] $-\kappa/2 < \xi < -\kappa/2\sqrt{3}$ and $\kappa/2\sqrt{3}
<\xi<\kappa/2$. In this case the external scale factor first contracts then
expands, bouncing from infinity to infinity. In particular, for 
\begin{itemize}
\item[a)] $-\kappa/2<\xi\le-\kappa/3$ ($p_\beta<-2\kappa$) the internal scale
factor shrinks from infinity to zero. The Hubble parameter starts with infinite
negative value, becomes positive and then decreases to zero after having
reached a positive maximum. $g$ and $g_{10}$ decrease from infinity to zero
[$g_{10}$ to a finite nonzero positive value for the limiting value
$\xi=-\kappa/3$];
\item[b)] $-\kappa/3<\xi\le-\kappa/2\sqrt{3}$ the internal scale factor shrinks
from infinity to zero [$-\kappa/3<\xi<-\kappa/2\sqrt{3}$
($-2\kappa<p_\beta<0$)] or is constant [$\xi=-\kappa/2\sqrt{3}$ 
($p_\beta=0$)]. The Hubble parameter starts with infinite negative value,
becomes positive and then decreases to zero after having reached a positive
maximum. $g$ decreases from infinity to zero. $g_{10}$ bounces from infinity
to infinity via a positive minimum;
\item[c)] $\kappa/2\sqrt{3}\le\xi<\kappa/3$ ($-2\kappa<p_\beta<0$). The
internal scale factor shrinks from infinity to zero
[$\kappa/2\sqrt{3}<\xi<\kappa/3$ ($-2\kappa<p_\beta<0$)] or is constant [$\xi
=\kappa/2\sqrt{3}$ ($p_\beta=0$)]. The Hubble parameter is first negative and
small, decreases to a negative minimum and then increases to infinity.  $g$
increases from zero to infinity.  $g_{10}$ bounces from infinity to infinity
via a positive minimum.
\item[d)] $\kappa/3\le\xi<\kappa/2$ ($p_\beta<-2\kappa$). The internal scale
factor shrinks from infinity to zero. The Hubble parameter is first negative
and small, decreases to a negative minimum and then increases to infinity. $g$
increases from zero to infinity. $g_{10}$ decreases from infinity to zero for
$\kappa/3<\xi<\kappa/2$ or to a finite nonzero positive value for the limiting
value $\xi=\kappa/3$.
\end{itemize}
\end{itemize}
Scenarios {\it i)} and {\it iii)} may be suitable candidates for a physical
description of a late time expanding universe emerging from a strong coupling
region. According to {\it i)} a decelerated universe begins in a strong
coupling region with large coupling constants, $g$ and $g_{10}$, and internal
dimensions much larger than the external dimensions. Though this might be seen
as a kind of severe fine-tuned initial conditions, for early times we are in
the strong coupling regime of the theory, where the spacetime curvature blows
up, and we expect the low-energy description of M-theory to break down.
Possibly, nonperturbative effects will cure initial conditions and provide a
mechanism for early inflation. Inflation happens in case {\it iii)}, where the
external spacetime is first contracting and eventually expanding, thus evolving
through an accelerated expanding phase. However, bouncing universes do not have
sufficient inflationary e-foldings to solve the horizon problem. Indeed, we
find
\be
{a_f\over a_i}=\left[{1\over 2}\left(\sqrt{p^{-2}-3}-1\right)\right]^{1/4}
\left[{2p^2-1+p\sqrt{p^{-2}-3}\over p(2p+1)}\right]^{-p/2}\,,
\ee
where $p=\xi/\kappa$, and $a_i$ and $a_f$ are the external scale factors at the
beginning and at the end of the inflationary phase, respectively. For
$-1/2<p<-1/(2\sqrt{3})$ and $1/(2\sqrt{3})<p<1/2$ the ratio $a_f/a_i$ is always
finite and $\approx\sqrt{2}$. 
\subsection{Quantization}
Turning to the hybrid canonical chart, from the constraint
(\ref{constraint-hybII}) it is natural to choose the operators $\hat p_a$,
$\hat p_b$, $\hat p_c$ and $\hat p_\sigma$ as
\be
\hat p_a=-i{\partial\over\partial a}\,,\quad
\hat p_b=-i{\partial\over\partial b}\,,\quad
\hat p_c=-i{\partial\over\partial c}\,,\quad
\hat p_\sigma=-i{\partial\over\partial\sigma}\,.
\label{op-hybII}
\ee
The Wheeler-De Witt (WDW) equation is
\be
\left[-{\partial^2~\over\partial a^2}
+{\partial^2~\over\partial b^2}
-{\partial^2~\over\partial c^2}
-e^{-2a}{\partial^2~\over\partial\sigma^2}\right]
\Psi(a,b,c,\sigma)=0\,.
\label{WDWII}
\ee
The WDW equation can be completely solved by the technique of separation of
variables. The general (bounded) solution is the superposition of wave
functions
$$
\Psi(a,b,c,\sigma)=\int dk_b dk_c dk_\sigma A(k_b,k_c,k_\sigma)
\psi(k_b,k_c,k_\sigma;a,b,c,\sigma)\,,
$$
\be
\psi(k_b,k_c,k_\sigma;a,b,c,\sigma)={\cal N} e^{\pm ib k_b}e^{\pm ic k_c}e^{\pm i\sigma k_\sigma}
K_{i\nu}(k_\sigma e^{-a})\,,\quad\nu=\sqrt{k_b^2-k_c^2}\,
\label{WDW-solII}
\ee
where $K_{i\nu}$ is the modified Bessel function of imaginary index $i\nu$. 
By properly choosing the normalization factor $\cal N$, and fixing the
gauge using the $b$ degree of freedom, the eigenstates of the physical
Hamiltonian with energy $E=k^2_b$ read
\be
\psi_{k_b,k_c,k_\sigma}=\sqrt{\nu\sinh\pi\nu\over 2\pi^4}
e^{\pm ic k_c}e^{\pm i\sigma k_\sigma} K_{i\nu}(k_\sigma e^{-a})\,.
\label{waves-hybII}
\ee

Let us briefly discuss the correspondence between the hybrid wave functions and
the classical solutions. The oscillating regions of the wave functions
correspond to the classically allowed regions of the configuration space. Along
$c$ and $\sigma$ directions the wave functions (\ref{waves-hybII}) are
described by plane waves. Along the $a$ direction the wave functions are
oscillating in the region
\be
0< e^{-a}\laq {\nu\over k_\sigma}\,.
\ee
This corresponds to the classically allowed region for the hybrid variable
$a$. (We have chosen $k_\sigma>0$ for simplicity.) Indeed, from the solutions
of the equations of motion we have
\be
0<e^{-a}={\kappa\over p_\sigma}[\cosh(\kappa\tau)]^{-1}\le
{\kappa\over p_\sigma}\,.
\ee
The wave functions go like $e^{\pm ia\nu}$ for large values of $a$. Finally,
the relation between the quantum numbers $k_i$ and the classical parameters
that characterize the behaviour of the classical solution is
\be
k_b=-2\sqrt{3}\xi\,,\qquad\qquad k_c={p_\beta\over 2\sqrt{3}}\,,\qquad\qquad\nu=\kappa\,.
\ee
\section{Conclusions\label{concl}}
In this talk we have analysed a simple spatially flat, four-dimensional
cosmological model derived from the M-theory effective action, Eq.\ (\ref{b1}).
The eleven-dimensional metric is first compactified on a one-dimensional circle
to obtain the type IIA superstring effective action and then on a six-torus to
obtain the effective four-dimensional theory. In our investigation we
concentrated the attention on the boundary of the physical phase space of the
theory, and in particular  to the invariant submanifold with negligible RR
four-form field strength. In our discussion we have heavily employed the
canonical formalism. This approach makes the analysis of the features of the
classical solution extremely simple and allows a straightforward quantization
of the theory.

In the classical setting, we have found regions in the moduli space where a
four-dimensional FRW universe evolves from a strong coupling regime towards a
weak coupling regime, both internal six-volume and eleven-dimension
contracting. The dynamics may also be characterized by an early accelerated
(inflationary) expansion with the spacetime eventually approaching a standard
FRW decelerated expansion. 

The quantization of the two invariant submanifolds can be performed exactly and
the Hilbert space of states can eventually be obtained. In the quantum
framework our analysis allows to identify the quantum states that correspond to
the different classical behaviours. In the hybrid representation we have
identified regions in the space of parameters where the wave function of the
universe is either oscillating or exponentially decaying. These regions are
determined by the inverse exponential function of the four-dimensional
(unshifted) dilaton, i.e., by the four-dimensional string coupling, and
correspond to classically allowed and classically forbidden regions,
respectively. Starting from the Hilbert space of states, the quantum mechanics
of M-theory cosmology can be constructed with aid of usual elementary quantum
mechanics techniques.

\section*{Acknowledgments}
We are very grateful to M.\ Gasperini, M.\ Henneaux, N.\ Kaloper, J.\ Lidsey,
C.\ Ungarelli, A.\ Vilenkin and D.\ Wands for interesting discussions and
useful comments. This work is supported by grants ESO/PROJ/12\-58/98,
CERN/P/FIS/15190/1999, Sapiens-Proj32327/99. M.C.\ is partially supported by
the FCT grant Praxis XXI BPD/20166/99. This work is supported in part by funds 
provided by  the U.S. Department of Energy (D.O.E.) under cooperative 
research agreement DE-FC02-94ER40818.


\begin{thebibliography}{999}
%
%
\bibitem{strings}{See e.g.\  M.\ Kaku, {\it Introduction to superstrings and
M-theory} (Springer Verlag, NY, 1999); M.B.\ Green, J.H.\ Schwarz and E.\
Witten, {\it Superstring theory}, Vol.\ I and II (Cambridge University Press,
Cambridge, 1987); J.\ Polchinski, {\it String theory}, Vol.\ I and II
(Cambridge University Press, Cambridge, 1998).}
%
\bibitem{prebb}{See http://www.to.infn.it/{\~\null}gasperin for an updated
collection of papers on pre-big bang cosmology; M.\ Gasperini,
\Journal{\CQG}{17}{R1}{2000} [{\tt hep-th/0004149}]; G.\ Veneziano, ``String
cosmology: the pre-big bang scenario'', Lectures given at 71st Les Houches
Summer School: The primordial universe, Les Houches, France, 28 June -23 July
1999 [{\tt hep-th/0002094}]; M.\ Gasperini and G.\ Veneziano,
\Journal{\ASP}{1}{317}{1993} [{\tt hep-th/9211021}]; J.\ Lidsey, D.\ Wands and
E.\ Copeland, \PRP\ (2000) to appear [{\tt hep-th/9909061}].}
%
%
\bibitem{exit}{M.\ Gasperini and G.\ Veneziano, \Journal{\GRG}{28}{1301}{1996}
[{\tt hep-th/9602096}]; M.\ Gasperini, J.\ Maharana and G.\ Veneziano,
\Journal{\NPB}{472}{349}{1996} [{\tt hep-th/9602087}]; M.\ Gasperini, M.\
Maggiore and G.\ Veneziano, \Journal{\NPB}{494}{315}{1997} [{\tt
hep-th/9611039}]; M.\ Cavagli\`a and C.\ Ungarelli,
\Journal{\CQG}{16}{1401}{1999} [{\tt gr-qc/9902004}]; R.\ Brustein and R.\
Madden, \Journal{\PRD}{57}{712}{1998} [{\tt hep-th/9708046}];
\Journal{\PLB}{410}{110}{1997} [{\tt hep-th/9702043}]; S.\ Foffa, M.\ Maggiore
and R.\ Sturani, \Journal{\NPB}{552}{395}{1999} [{\tt hep-th/9903008}].}
%
\bibitem{LOW}{A.\ Lukas, B.A.\ Ovrut and D.\ Waldram,
\Journal{\PRD}{60}{086001}{1999}; [{\tt hep\--th/9806022}];
\Journal{\PRD}{61}{023506}{2000} [{\tt hep-th/9902071}];
\Journal{\NPB}{495}{365}{1997} [{\tt hep-th/9610238}]; A.\ Lukas and B.A.\ Ovrut,
\Journal{\PLB}{437}{291}{1998} [{\tt hep-th/ 9709030}]; A.\ Lukas, B.A.\ Ovrut,
K.S.\ Stelle and D.\ Waldram, \Journal{\PRD}{59}{086001}{1999} [{\tt
hep-th/9803235}].}
%
\bibitem{RS}{L.\ Randall and R.\ Sundrum, \Journal{\PRL}{83}{3370}{1999} [{\tt
hep-th/9905221}]; ibid.\ 4690 (1983) [{\tt hep-th/9906064}].}
%
\bibitem{HW}{E.\ Witten, \Journal{\NPB}{471}{135}{1996} [{\tt hep-th/9602070}];
P.\ Ho{\v r}ava and E.\ Witten, \Journal{\NPB}{460}{506}{1996} [{\tt
hep-th/9510209}]; ibid.\ 475, 94 (1996) [{\tt hep-th/9603142}].}
%
\bibitem{DH}{T.\ Damour and M.\ Henneaux, \Journal{\PRL}{85}{920}{2000} [{\tt
hep-th/0003139}]; \Journal{\PLB}{488}{108}{2000} [{\tt hep-th/0006171}].}
%
\bibitem{LMMP}{H.\ Lu, J.\ Maharana, S.\ Mukherji and C.N.\ Pope,
\Journal{\PRD}{57}{2219}{1998} [{\tt hep-th/9707182}].}
%
%
\bibitem{CLW}{E.J.\ Copeland, A.\ Lahiri and D.\ Wands,
\Journal{\PRD}{50}{4868}{1994} [{\tt hep\--th/9406216}].}
%
\bibitem{KKO}{N.\ Kaloper, I.I.\ Kogan and K.A.\ Olive,
\Journal{\PRD}{57}{7340}{1998} and Erratum, ibid.\ 60, 049901 (1999) [{\tt
hep-th/9711027}].}
%
\bibitem{BCLN}{A.P.\ Billyard, A.A.\ Coley, J.E.\ Lidsey, U.S.\ Nilsson,
\Journal{\PRD}{61}{043504}{2000} [{\tt hep-th/9908102}].}
%
\bibitem{MC-PM}{M.\ Cavagli\`a, P. Moniz, ``Canonical and quantum FRW
cosmological solutions in M-theory'', \CQG\ in press [{\tt hep-th/0010280}].}

\end{thebibliography}
\end{document}